\documentclass[twocolumn]{aastex631}
\usepackage{apjfonts}
\usepackage{graphicx}
\usepackage{amsmath}
\usepackage{amssymb}
\usepackage{color}

\gdef\xx[#1]{\textcolor{red}{#1}}

\gdef\kms{km\,s$^{-1}$}
\gdef\msun{M$_{\odot}$}

\gdef\lya{Ly\kern 0.09em$\alpha$}
\gdef\ha{H\kern 0.09em$\alpha$}

\newcommand{\GG}[1]{}
\lefthead{van Dokkum \& Conroy}
\righthead{}
\begin{document}

\newcommand\XXX[1]{{\textcolor{red}{\textbf{x\ #1\ x}}}}

\title{Reconciling $M/L$ Ratios Across Cosmic Time: a Concordance IMF for Massive Galaxies}

%\correspondingauthor{Pieter van Dokkum}

\author[0000-0002-8282-9888]{Pieter van Dokkum}
\affiliation{Astronomy Department, Yale University, 219 Prospect St,
New Haven, CT 06511, USA}
\author[0000-0002-1590-8551]{Charlie Conroy}
\affiliation{Harvard-Smithsonian Center for Astrophysics, 60 Garden Street,
Cambridge, MA, USA}

\begin{abstract}

  The stellar initial mass function (IMF) is thought to be
  bottom-heavy in the cores of the most massive galaxies, with an
  excess of low mass stars compared to the Milky Way.  However,
  studies of the kinematics of quiescent galaxies at $2<z<5$ find
  $M/L$ ratios that indicate lighter IMFs.  Light
  IMFs have also been proposed for the unexpected populations of
  luminous galaxies that JWST has uncovered at $z>7$, to reduce
  tensions with galaxy formation models.  Here we explore `ski slope'
  IMFs that
  are simultaneously bottom-heavy, with a steep slope at low stellar
  masses, and top-heavy, with a shallow slope at high masses.  We
  derive a form of the IMF for massive galaxies that is consistent
  with measurements in the local universe and yet produces relatively
  low $M/L$ ratios at high redshift. This concordance IMF has slopes
  $\gamma_1=2.40\pm0.09$, $\gamma_2=2.00\pm0.14$, and
  $\gamma_3=1.85\pm0.11$ in the regimes 0.08\,\msun\,--\,0.5\,\msun,
  0.5\,\msun\,--\,1\,\msun, and $>1$\,\msun\ respectively. The IMF
  parameter $\alpha$, the mass excess compared to a Milky Way IMF,
  ranges from $\log(\alpha)\approx+0.3$ for present-day galaxies to
  $\log(\alpha)\approx-0.1$ for their star forming progenitors.  The
  concordance IMF applies only to the central regions of the most
  massive galaxies, with velocity dispersions $\sigma\sim300$\,\kms,
  and their progenitors. However, it can be generalized
 using a previously-measured relation between
  $\alpha$ and $\sigma$. We arrive at the
  following modification to the Kroupa (2001) IMF for galaxies with
  $\sigma\gtrsim 160$\,\kms:
  $\gamma_1\approx1.3+4.3\log\sigma_{160}$;
  $\gamma_2\approx2.3-1.2\log\sigma_{160}$; and
  $\gamma_3\approx2.3-1.7\log\sigma_{160}$, with
  $\sigma_{160}=\sigma/160$\,\kms.  If galaxies grow primarily
  inside-out, so that velocity dispersions are relatively stable,
  these relations should also hold at high redshift.

\end{abstract}

%\keywords{
%galaxies: evolution --- galaxies: structure }

\section{Introduction}

The form of the stellar initial mass function (IMF), and its potential
variation with star forming conditions, is a topic of ongoing debate
\citep[see the reviews by][]{bastian:10,offner:14,smith:20}.  The
question has taken on a new significance since the launch of JWST, as
unexpected populations of bright and seemingly massive galaxies have
been discovered at $z>7$
\citep[e.g.,][]{naidu:22,labbe:23,xiao:23}. The formation of the
stellar component of these galaxies requires extremely rapid
conversion of gas into stars, perhaps more rapid than can be
accommodated in standard galaxy formation models
\citep[see][]{boylankolchin:23,dekel:23}.  There is a straightforward
way to reduce or even eliminate this tension, as the total mass of a
stellar population is dominated by stars that have low masses and
contribute very little to the light. Therefore, as pointed out by
\citet{steinhardt:23} and others \citep{woodrum:23,wang:24}, if the
IMF in these galaxies were top-heavy (or bottom-light) their derived
masses would be lowered considerably.  Galaxy formation models with
top-heavy IMFs also provide better fits to the UV luminosity function
at these redshifts \citep{inayoshi:22,harikane:23,yung:24,schaerer:24}.

There is good evidence that such IMFs can occur in certain
environments; star clusters in the central
$\sim 300$\,pc of the Milky Way appear to have an excess of high mass
stars compared to young clusters elsewhere in the disk
\citep{lu:13,hosek:19,chabrier:24}, as does
the 30 Doradus cluster in the Large Magellanic Cloud
\citep{schneider:18}. Top-heavy IMFs for early galaxies
have also been proposed on theoretical grounds, as higher CMB
temperatures can lead to an increase in the turnover mass
\citep{larson:98,bate:23}.  However, the IMF in the central regions of
massive early-type galaxies appears to be {\em bottom}-heavy, with an
excess of low mass stars compared to the Milky Way \citep[see][for a
review]{smith:20}. The evidence for bottom-heavy IMFs comes primarily
from studies of gravity-sensitive stellar absorption lines
\citep{cenarro:03,dokkum:10,spiniello:12,conroy:imf12,labarbera:13,martinnavarro:15,
  labarbera:19}, which detect the low mass stars directly, and also
from higher-than-expected stellar $M/L$ ratios derived from
gravitational lensing \citep{treu:10,posacki:15} and galaxy dynamics
\citep{cappellari:12,li:17,shetty:20}.  As the old and dense central
regions of massive galaxies in the present-day Universe are the likely
descendants of the most luminous $z>7$ galaxies, these results suggest
that -- if anything -- the masses of the JWST-discovered galaxies
might be {\em under}estimated, by factors of 1.5--3.

In this context, massive quiescent galaxies at $z\approx2-5$ are an
important bridge population. As they lack dust and young stars their
stellar masses are usually well-constrained (for an assumed IMF), and
dynamical masses can be determined from their sizes and absorption
line kinematics. The ratio of dynamical mass to stellar mass then
constrains the form of the IMF, particularly if this ratio is close to
unity: a high ratio could be attributed to some combination of dark
matter, gas, and extra stellar mass, but a low ratio is only
consistent with relatively light IMFs.  While one study reported
evidence for increased $M_{\rm dyn}/M_*$ ratios at $z\sim 2$
\citep{belli:17}, most find the opposite effect, that is, a decrease
compared to local values \citep{sande:13,mendel:20,kriek:24}.  As
discussed in \citet{kriek:24}, the $M/L$ ratios of quiescent galaxies
at $z\sim 2$ appear to be in tension with the bottom-heavy IMFs that
have been found for similar-mass galaxies at $z=0$, when systematic
effects are accounted for.
Dynamical mass measurements at $z>2$ currently have large
uncertainties, but they appear to be consistent with this trend:
\citet{esdaile:21} and \citet{carnall:24} find
$M_{\rm dyn}/M_* \sim 1$ for small samples of massive quiescent
galaxies at $3< z < 5$, when assuming a standard Kroupa or Chabrier
IMF.  Taken together, these results suggest a possible gradual
evolution, with the youngest (generally highest redshift) massive
quiescent galaxies having stellar $M/L$ ratios consistent with a
standard IMF, and the oldest (lowest redshift) galaxies having $M/L$
ratios that indicate extra mass.

One way to reconcile these seemingly contradictory results is to
appeal to complex formation histories: it may be that the stars we see
in massive galaxies at $z\sim 2$ and above are only a small fraction
of those in the cores of present-day ellipticals. In that case the IMF
at high redshift could be quite different from the one that produced
most of the stars that we see today.  Massive galaxies certainly grow
substantially in size and mass throughout their history, through a
combination of star formation and mergers \citep[see,
e.g.,][]{dekel:09mass}.  However, the already-high central densities
of early massive galaxies \citep[even out to $z>7$;
see][]{bezanson:09,baggen:23}, as well as the old ages of the cores of
ellipticals \citep[e.g.,][]{greene:15}, both strongly suggest that
most of this activity happened outside of the central regions. One
viable form of complex evolution that should be mentioned is a mode of
late central star formation that only produces low mass stars, as has
been hypothesized to occur in cooling flows \citep[see][]{fabian:24}.

In this paper we take a different approach. We assume that the most
massive galaxies at high redshift are representative of the
progenitors of the central regions of the most massive galaxies at
lower redshifts, and that the behavior of the $M/L$ ratios reflects a
particular form of the IMF.  We use the observed $M/L$ ratios (of
living stars and remnants, and of living stars only) to constrain this
form. This approach has its roots in the work of \citet{tinsley:80},
who first pointed out that the form of the IMF near the turnoff mass
could be determined from the luminosity evolution of galaxies.

\section{Methodology}

We systematically explore
a family of IMF shapes, and ask what particular form in that family produces the lowest
$M/L$ ratios at high redshift while being consistent with the bottom-heavy IMFs
that have been derived for the cores of local ellipticals.

\subsection{General Functional Form}

\begin{figure}[t!]
  \begin{center}
  \includegraphics[width=0.95\linewidth]{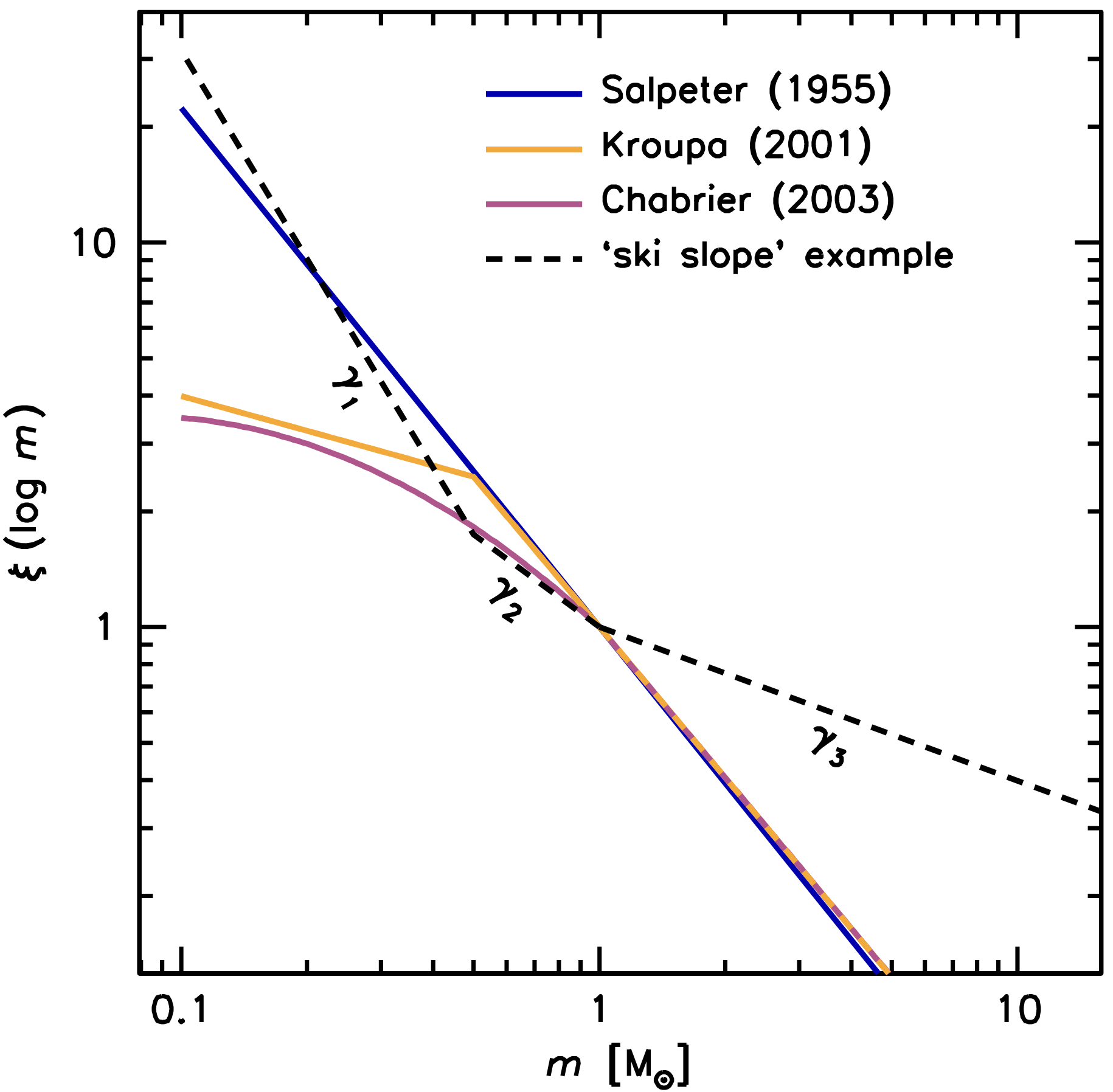}
  \end{center}
\vspace{-0.2cm}
    \caption{
Illustration of the general form of the IMFs that are explored in this paper: a steep, Salpeter-like
slope at low masses, followed by a flattening between $0.5$\,\msun\ and $1$\,\msun, and
a further flattening at $>1$\,\msun. This gradual flattening
somewhat resembles a ski slope. 
For context and reference, the widely used \citet{salpeter:55}, \citet{kroupa:01}, and
\citet{chabrier:03} forms are also shown.
Ski slope IMFs
are simultaneously bottom-heavy and top-heavy compared to the
IMF of the Milky Way.
}
\label{intro.fig}
\end{figure}

The IMFs are parameterized by the three power law segments that are
familiar from the \cite{kroupa:01} IMF: 0.08\,\msun\,$<m<0.5$\,\msun,
0.5\,\msun\,$<m<1$\,\msun, and $m>1$\,\msun. The power law slopes in
these regions are denoted with $\gamma_1$, $\gamma_2$, and $\gamma_3$
respectively.  The \citet{kroupa:01} IMF, appropriate for star
formation in the Milky Way disk, has $\gamma_1=1.3$, $\gamma_2=2.3$,
and $\gamma_3=2.3$. It is shown in Fig.\ \ref{intro.fig}, along with
the (functionally nearly identical) \citet{chabrier:03} IMF and the
\citet{salpeter:55} IMF (which has
$\gamma_1 = \gamma_2 = \gamma_3=2.35$).  We use integration limits of
$0.08$\,\msun\,--\,$100$\,\msun.

Within this general framework, the family of IMFs that we explore have the following characteristics:
\begin{enumerate}
  \item{A steep slope at low masses, $2.3\leq \gamma_1\leq 2.7$. This is needed to reproduce
  the spectroscopic evidence for low mass stars reported by \citet{conroy:imf12} and many others.
   The upper end of the range reflects the most bottom-heavy IMF that has been found so far,
   in the heart of the giant elliptical galaxy NGC\,1407: using a non-parametric form of the IMF,
\citet{conroy:17} find a slope of $2.7$ all the way to the hydrogen burning limit. Such a steep
slope has also been derived theoretically for progenitors of massive ellipticals \citep{chabrier:14}.}
\item{A shallower slope at intermediate masses, that is at most as steep as the Salpeter value:
$\gamma_2\leq \gamma_1$ and $\gamma_2\leq 2.3$. Whereas $\gamma_1$ controls the amount
of mass in (living) stars, $\gamma_2$ controls the luminosity: $\sim 95$\,\% of the luminosity
of an old stellar population comes from stars in the range 0.5\,\msun\,$<m<$\,1\,\msun\ \citep[see,
e.g., Fig.\ 2 in][]{conroy:12}. We also impose the criterion
$\gamma_1 - \gamma_2 \leq 1.0$; as we will see later, no viable solutions come near this limit.
}
\item{An even shallower slope at high masses, again at most as steep as Salpeter:
$\gamma_3\leq\gamma_2$ and $\gamma_3\leq 2.3$. We also
require $\gamma_2 - \gamma_3\leq 1.0$.
The high mass slope has no
impact on the luminosity of the cores of ellipticals, as there are no
living stars with $m>1$\,\msun. Instead, it controls the luminosity of the stellar
population at younger ages and higher redshifts (together with $\gamma_2$), and the amount
of mass at late times that is locked up in remnants (neutron stars and black holes).
Remnants are a small fraction of the total mass for a Salpeter IMF, but they become increasingly
important for more bottom-light and/or top-heavy IMFs.
As noted by \citet{maraston:98} and others, the total mass of a stellar population with
a top-heavy IMF is often remnant-dominated at late times.
}
\end{enumerate}

The general shape of the IMFs that we explore is illustrated with the broken line in Fig.\ \ref{intro.fig}.
The limiting case is a near-Salpeter IMF, with $\gamma_1 = \gamma_2 = \gamma_3 = 2.3$,
but the general form is
{\em both} bottom-heavy and top-heavy. 

\subsection{Observational Constraints}
\label{sec:constraints}

We focus on three key aspects when assessing the viability of a particular IMF:
the number of low mass stars for old ages,
which has to match the constraints from absorption line
studies; the total $M/L$ ratio for old ages,
which has to match constraints from lensing and dynamics;
and the total $M/L$ ratio for young ages, which should be as low as possible
while still satisfying the first two constraints.

For convenience, and to allow straightforward comparisons to the literature, the
diagnostic parameters are expressed in terms of the mass excess compared to the
standard \citet{kroupa:01} IMF:
\begin{equation}
\alpha \equiv \frac{(M/L)[\gamma_1,\gamma_2,\gamma_3]}{(M/L)[1.3,2.3,2.3]}.
\end{equation}
Luminosities are calculated in the SDSS\,$r$ band. Masses can refer to the
total stellar mass, that is, living stars and remnants ($\alpha_{\rm tot}$),
or to the mass in living stars only ($\alpha_{\rm liv}$).
As $\sim 70$\,\% of the total living mass is locked up in stars with
masses $<0.5$\,\msun, $\alpha_{\rm liv}$ is a proxy for the ratio of
the number of low mass stars to the number of turnoff stars.

\begin{figure*}[tbp]
  \begin{center}
  \includegraphics[width=0.95\linewidth]{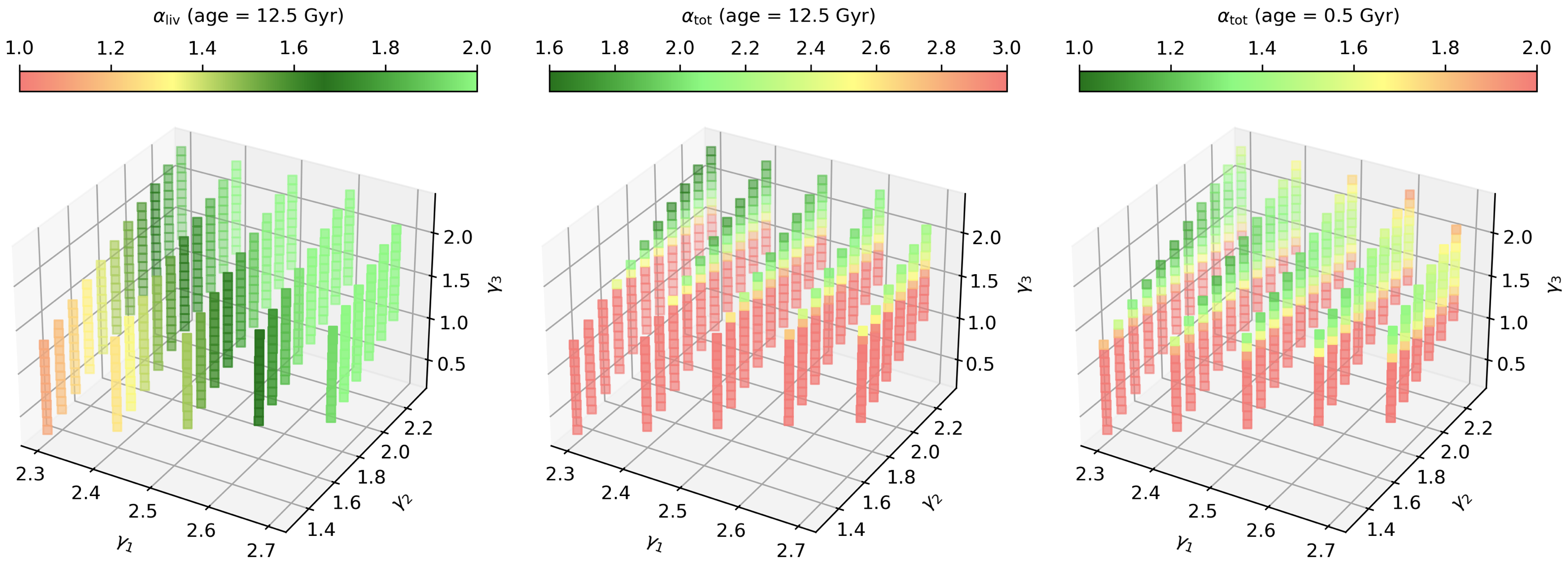}
  \end{center}
\vspace{-0.2cm}
    \caption{
Values of the IMF parameter $\alpha$ for all 484 `ski slope' IMFs that are considered
in this study, as a function of the three power law slopes $\gamma_1$, $\gamma_2$, $\gamma_3$
that define the shape. In each panel the color palette is such that red is
disfavored/disallowed and green is favored/allowed.
{\em Left:} $\alpha_{\rm liv}$, the mass excess
in living stars, for an age of 12.5 Gyr. Many IMFs satisfy the constraint $\alpha_{\rm liv}>1.6$.
{\em Middle:} $\alpha_{\rm tot}$, the total mass excess, for an age of 12.5 Gyr. This is a strong function of
the high mass slope, $\gamma_3$. Shallow high mass slopes produce too many remnants,
and are disallowed by the criterion $\alpha_{\rm tot}<2.3$. {\em Right:} $\alpha_{\rm tot}$
for an age of 0.5\,Gyr. The lowest values of $\alpha_{\rm tot}$ are in the range
$1.1-1.3$.
}
\label{alpha_constraints.fig}
\end{figure*}

With these definitions in place, we impose the following quantitative
requirements on viable IMFs:
\begin{eqnarray}
\alpha_{\rm liv} &>& 1.6\hspace{0.5cm}{\rm (age}\,=12.5\,\,{\rm Gyr)}\\
\alpha_{\rm tot} &<& 2.3 \hspace{0.5cm}{\rm (age}\,=12.5\,\,{\rm Gyr)}\\
\alpha_{\rm tot} &<& 1.3 \hspace{0.5cm}{\rm (age}\,=0.5\,\,{\rm Gyr)}
\end{eqnarray}
The first constraint ensures that there is an excess of low mass stars
compared to the Milky Way IMF. It is based on the spectral analysis 
of \citet{conroy:imf12},
who find $1.6\lesssim \alpha_{\rm tot}\lesssim 2.0$ for the most massive
and Mg-enhanced galaxies. Note that
\citet{conroy:imf12} express their results in terms of
$\alpha_{\rm tot}$ rather than $\alpha_{\rm liv}$. For
$\gamma_3=2.3$, as is assumed in that study,  $\alpha_{\rm liv}\approx 1.1\alpha_{\rm tot}$.
The second constraint is to ensure that the total mass
does not exceed constraints
from lensing and dynamics. \citet{posacki:15} find 
$\alpha_{\rm tot}\approx 1.6$ for a sample of strong lenses
with $\sigma \sim 300$\,\kms,
and \cite{mendel:20} find $\alpha_{\rm tot} \approx 2.2$ for SDSS galaxies
in the same mass range. A limit is needed: as we show later, 
IMFs with shallow high mass slopes (i.e., low
$\gamma_3$) are often remnant-dominated, yielding very high values
of $\alpha_{\rm tot}$.

The third constraint is of a different nature than the other two. As explained in
the Introduction, our aim is to determine whether it is possible to have low values
of $\alpha$ at high redshift while satisfying the observational constraints at
low redshift. The constraint $\alpha_{\rm tot}<1.3$ for young ages simply selects
the IMFs that produce the lowest $M/L$ ratios in this family of IMFs.

\subsection{Grid Search}

We perform a straightforward grid search to determine which IMFs satisfy the observational
constraints. The grid consists of all IMFs that
satisfy the constraints on $\gamma_1$, $\gamma_2$, and $\gamma_3$ that
were laid out in \S\,2.1, with steps of 0.1 in each of the three
parameters. Masses and luminosities are determined with the Python
implementation\footnote{Python-FSPS,
  https://dfm.io/python-fsps/current/} of the flexible stellar
population synthesis (FSPS) suite \citep{conroy:09,conroygunn:10}, for
dust-free single stellar populations with solar metallicity. The
default MIST isochrones \citep{choi:16} and MILES spectral library
\citep{miles} are used.

For each stellar population we determine the mass in living stars, the total
mass (in living stars and stellar remnants), and the luminosity in the SDSS
$r$ band, for an age of 0.5\,Gyr and an age of 12.5\,Gyr. The young age corresponds to
the approximate luminosity-weighted ages of the $3<z<5$ galaxies in \citet{esdaile:21} and
\citet{carnall:24}.
An age of 12.5 Gyr is appropriate for the cores of massive galaxies \citep[see, e.g.,][]{dokkum:17imf},
and also consistent with the age of the $z\sim 4$ galaxies if they evolve
passively to the present.
%Next, $M_{\rm tot}/L_r$ and $M_{\rm liv}/L_r$ ratios are determined for each IMF, and divided by
%the $M_{\rm tot}/L_r$ and $M_{\rm liv}/L_r$ ratios for a standard \citet{kroupa:01} IMF with
%{\tt imf1}\,=\,1.3, {\tt imf2}\,=\,2.3, and {\tt imf3}\,=\,2.3 (see \S\,)
With the masses and luminosities for each IMF at both ages, as well as the equivalent
values for a standard \citet{kroupa:01} IMF, the diagnostics $\alpha_{\rm liv}$, $\alpha_{\rm tot}$
(at 12.5\,Gyr) and $\alpha_{\rm tot}$ (at 0.5\,Gyr) are calculated.

\section{Results}

\subsection{Allowed IMF Shapes}

The results for all 484 IMFs are visualized in Fig.\ \ref{alpha_constraints.fig}. 
The left and middle panels show $\alpha_{\rm liv}$ and $\alpha_{\rm tot}$, respectively, for
an age of 12.5\,Gyr. The right panel shows $\alpha_{\rm tot}$ for 0.5\,Gyr. In each
panel the color coding is
chosen such that green indicates `allowed' (or `preferred'), and red indicates `not allowed'.

Turning first to the results for $\alpha_{\rm liv}$ at 12.5\,Gyr (left panel), we find that the constraint
$\alpha_{\rm liv}>1.6$ is satisfied for a wide range of IMF shapes. That is, most of the IMFs in this family
have a relatively high number of low mass stars. This is a direct consequence of the
chosen range of $\gamma_1$, between Salpeter (2.3) and steeper than Salpeter (2.7). At
fixed $\alpha_{\rm liv}$ there is an interplay
between $\gamma_1$ and $\gamma_2$, such that the ratio of the number of low mass stars to the number
of turnoff stars is conserved. Specifically, for $1.7<\alpha_{\rm liv}<1.9$ we find $\gamma_2\approx 5.9-1.6\,\times\gamma_1$. Note that the results for $\alpha_{\rm liv}$ are entirely independent of $\gamma_3$, the slope
of the IMF beyond the turnoff mass.

Turning to the results for $\alpha_{\rm tot}$ at 12.5\,Gyr, we find that for most of the IMFs
the value of $\alpha_{\rm tot}$ is too high: the constraint
$\alpha_{\rm tot}<2.3$ is satisfied only in a small region of parameter space.
The key parameter is $\gamma_3$; shallow high mass slopes produce too many
stellar remnants, leading to $M/L$ ratios that are higher than is allowed by
lensing and dynamical measurements of massive $z=0$ galaxies. To illustrate the strong
dependence on $\gamma_3$, all IMFs with $\gamma_3<1.5$ have $\alpha_{\rm tot}>3$, and all
IMFs with $\gamma_3>2$ have $\alpha_{\rm tot}<2.3$. 

The right panel of Fig.\ \ref{alpha_constraints.fig} shows $\alpha_{\rm tot}$ for young ages. The results
are superficially similar to the 12.5\,Gyr panel, but the scale is different: whereas the lowest value
of $\alpha_{\rm tot}$ at 12.5\,Gyr is 1.62, it is only 1.13 at 0.5\,Gyr -- that is, very close
to a standard Chabrier or Kroupa IMF. Importantly, the green region of parameter space,
indicating the lowest values of $\alpha_{\rm tot}$, has considerable overlap with the green regions
in the other two panels. This means that there are viable IMF solutions that satisfy all the
constraints of \S\,2.2.

\subsection{Concordance IMF}

\begin{figure}[tbp]
  \begin{center}
  \includegraphics[width=0.95\linewidth]{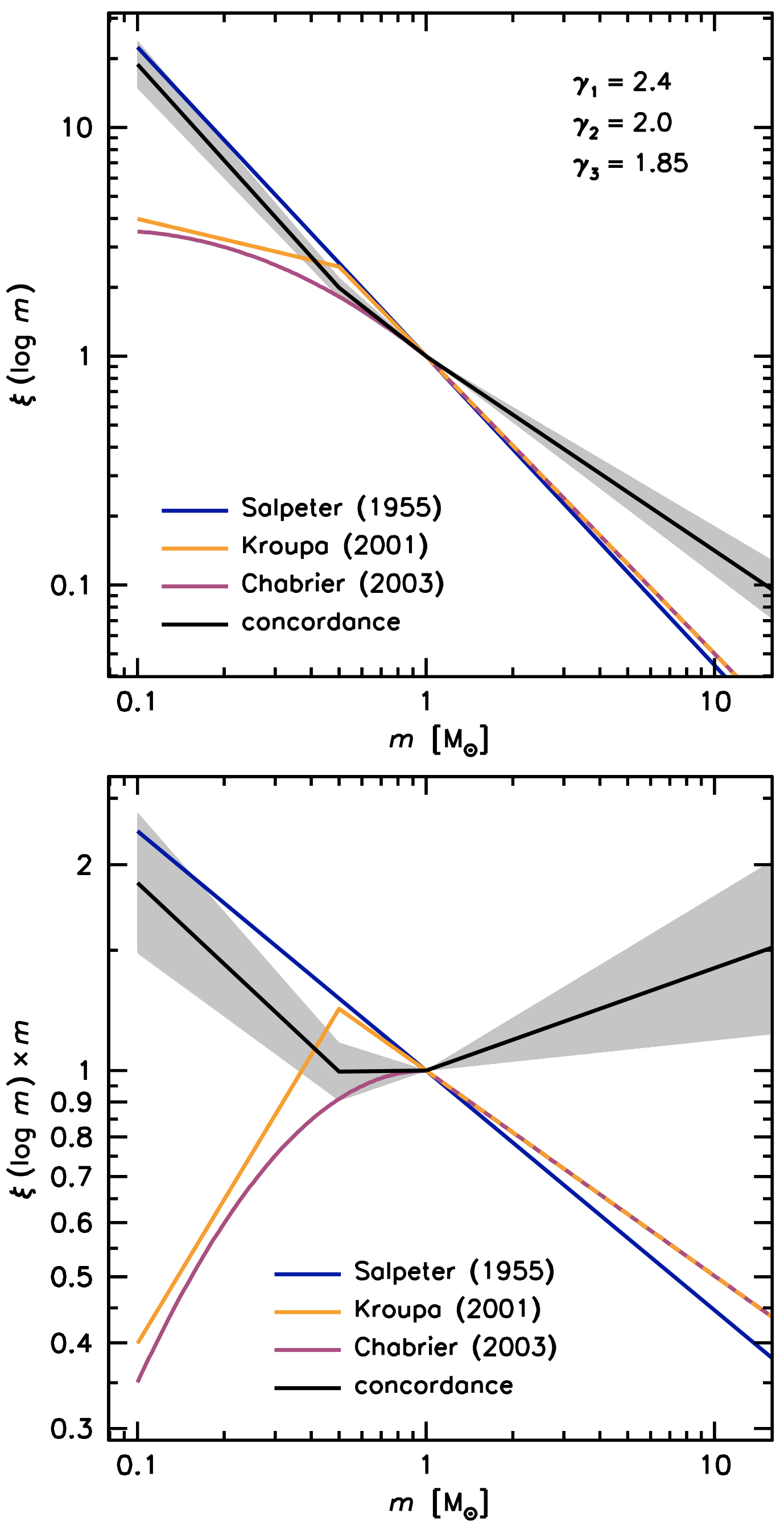}
  \end{center}
\vspace{-0.2cm}
    \caption{
The `concordance' IMF that we propose here for massive galaxies,
compared to standard forms. The same information is shown in the
top and bottom panels, but in the bottom panel the $y$-axis is multiplied by $m$ to reduce the
dynamic range and bring out the differences
between the IMFs. The turnover at 1\,\msun\ is not significant; IMFs with $\gamma_1 = 2.4$ and
$\gamma_2 = \gamma_3 = 1.9$ produce very similar results.
}
\label{concord.fig}
\end{figure}

Applying the criteria of Eqs.\ 2, 3, and 4 yields 20 viable IMFs that are all very similar. 
The mean power law slopes are
\begin{eqnarray}
\gamma_1 &=& 2.40\pm 0.09;\nonumber\\
\gamma_2 &=& 2.00\pm 0.14;\nonumber\\
\gamma_3 &=& 1.85\pm 0.11,\nonumber
\end{eqnarray}
where the uncertainty is taken to be the rms range among the 20 IMFs.
This concordance IMF is shown in Fig.\ \ref{concord.fig}. It is similar to the Salpeter
IMF in terms of its ratio of low mass stars to turnoff stars, but flatter at high masses.
The key turnover is at 0.5\,\msun: the slope changes at this point in all 20 viable IMFs, by
$0.4 \pm 0.2$.
The turnover at 1\,\msun\ is not significant; of the 20 viable models, 7 have $\gamma_2 = \gamma_3$.

\begin{figure*}[tbp]
  \begin{center}
  \includegraphics[width=0.95\linewidth]{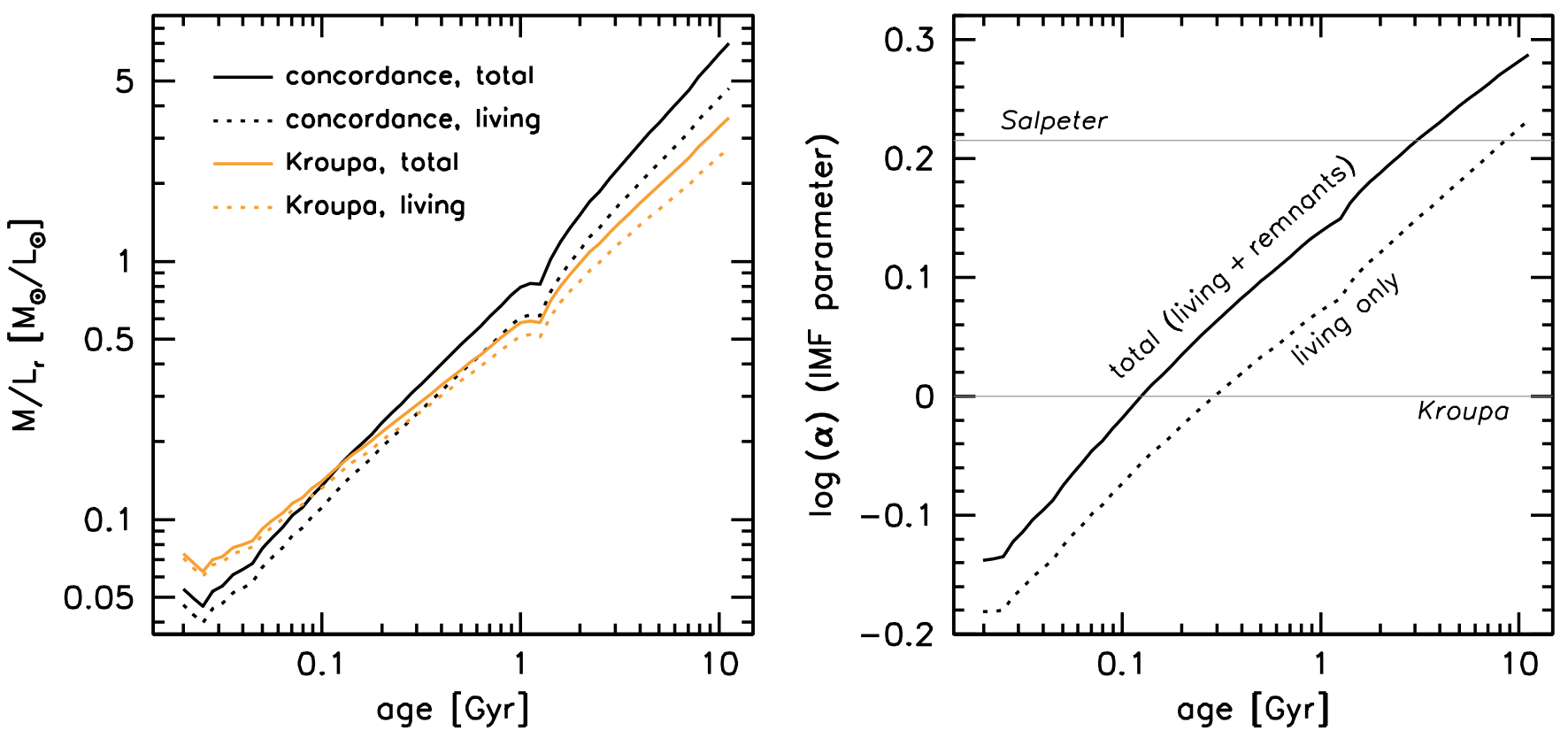}
  \end{center}
\vspace{-0.2cm}
    \caption{
{\em Left panel:} Relation between $M/L_r$ ratio and age, for the concordance IMF and for a standard
\citet{kroupa:01} IMF. For old ages the concordance $M/L$ ratios are higher than the Kroupa
$M/L$ ratios, but this reverses for the youngest ages. {\em Right panel:} IMF parameter $\alpha$
as a function of age, for the concordance IMF. This relation is the logarithmic offset between
the two curves in the left panel. While the mass excess is larger than that of a Salpeter IMF
for old ages, it is close to a Kroupa or Chabrier IMF for ages of 0.1\,--\,0.5\,Gyr. For the
youngest ages there is a mass deficit with respect to the Kroupa IMF.
}
\label{ml_age.fig}
\end{figure*}

The IMF is remarkably well constrained, both at the low mass end and at the high mass end.
As explained above, the three limits of Eqs.\ 2, 3, and 4 all control different aspects of the IMF.
The low mass slope is tightly constrained by the observed mass excess in low mass stars. The high
mass slope is constrained by two competing effects: if it
were shallower, stellar remnants would violate lensing and dynamical constraints for old ages,
and if it were steeper, the $M/L$ ratios for young ages would be too high.

\subsection{M/L Ratios Over Cosmic Time}

Returning to the questions that were asked in the Introduction, we now
consider how the IMF parameter $\alpha$ (the mass excess compared to a
Kroupa IMF) changes with redshift. The time evolution of the $M/L_r$
ratio is shown in the left panel of Fig.\ \ref{ml_age.fig}, for both
the concordance IMF and a standard Kroupa IMF.  In the right panel we
show the time evolution of $\alpha$, the ratio of these two
curves. The IMF parameter monotonically increases with age, going from
$\lesssim -0.1$ at ages of $\lesssim 50$\,Myr to $\approx +0.3$ at
12.5\,Gyr.

Turning to the galaxy populations that were discussed in the
Introduction, the concordance IMF reproduces the large mass excess,
and the large number of low mass stars, observed in the cores of
present-day massive elliptical galaxies; reproduces the smaller excess
of $0-0.2$\,dex that has been found for young quiescent galaxies at
$z=2-5$; and produces a mass {\em deficit} for massive star forming
$z>7$ galaxies, in better agreement with galaxy formation models.

\subsection{Dependence on the Low Mass Cutoff}

In our analysis the lower integration limit of the IMF is
the hydrogen burning limit of 0.08\,\msun\ \citep{chabrier:23}, that is,
it is assumed that the mass locked up in brown dwarfs is negligible.
For a Milky Way IMF the precise lower limit is not important
as most of the stellar mass is in stars near the turnover; for
a Kroupa IMF, changing the lower limit to 0.03\,\msun\ increases the total mass by 2\,\% and changing it to 0.01\,\msun\
increases the mass by 3\,\%. However, for a Salpeter IMF
the total mass is
very sensitive to the low mass cutoff, increasing
by a factor of 1.4 going from $0.08$\,\msun\ to $0.03$\,\msun.
For a lower limit of 0.01\,\msun\ the increase is a factor of 2.2,
which means that most of the stellar mass is in the form of brown dwarfs.

As ski slope IMFs have a Salpeter-like low mass slope, their total mass
is also very sensitive to the low mass cutoff. We repeated the analysis
for integration limits of 0.03\,\msun\ and 0.01\,\msun, and find that
{\em no} IMFs satisfy the critera of \S\,\ref{sec:constraints}. 
Relaxing the criteria  to
$\alpha_{\rm tot}<2.5$ (age\,=\,12.5\,Gyr) and
$\alpha_{\rm tot}<1.5$ (age\,=\,0.5\,Gyr), there is one solution
for a cutoff of 0.03\,\msun, with
$\gamma_1 = 2.3$, $\gamma_2 = 2.1$, and $\gamma_3 = 1.8$,
and still no solution for a cutoff of 0.01\,\msun.

We infer that a turnover near the hydrogen burning limit
is required in the context of ski slope IMFs, as otherwise the
total $M/L$ ratios would exceed constraints from dynamics and lensing.
The steep slope at low masses and the sharp cutoff at $0.08$\,\msun\
should be viewed as a simplified parameterization of a shift
in the turnover mass from
$\sim 0.3$\,\msun\ for the Milky Way to $\sim 0.1$\,\msun\ for
the concordance IMF. The actual form of the low mass IMF may be
similar to the IMFs explored in \citet{chabrier:14}, which combine
slopes of $2.3-2.7$ with a turnover near $0.1$\,\msun.

\subsection{Dependence on Velocity Dispersion}

So far the focus has been on a single form of the IMF, but the IMF
parameter $\alpha$ correlates with mass, velocity dispersion,
metallicity, and (within a galaxy) radius \citep[see,
e.g.,][]{conroy:imf12,martinnavarro:15,mendel:20}.  Our analysis can
be extended by taking this variation into account, thereby enabling a
compact parameterization of the full range of IMF shapes from Milky
Way-like galaxies to the cores of massive ellipticals.

\begin{figure}[t!]
  \begin{center}
  \includegraphics[width=1.0\linewidth]{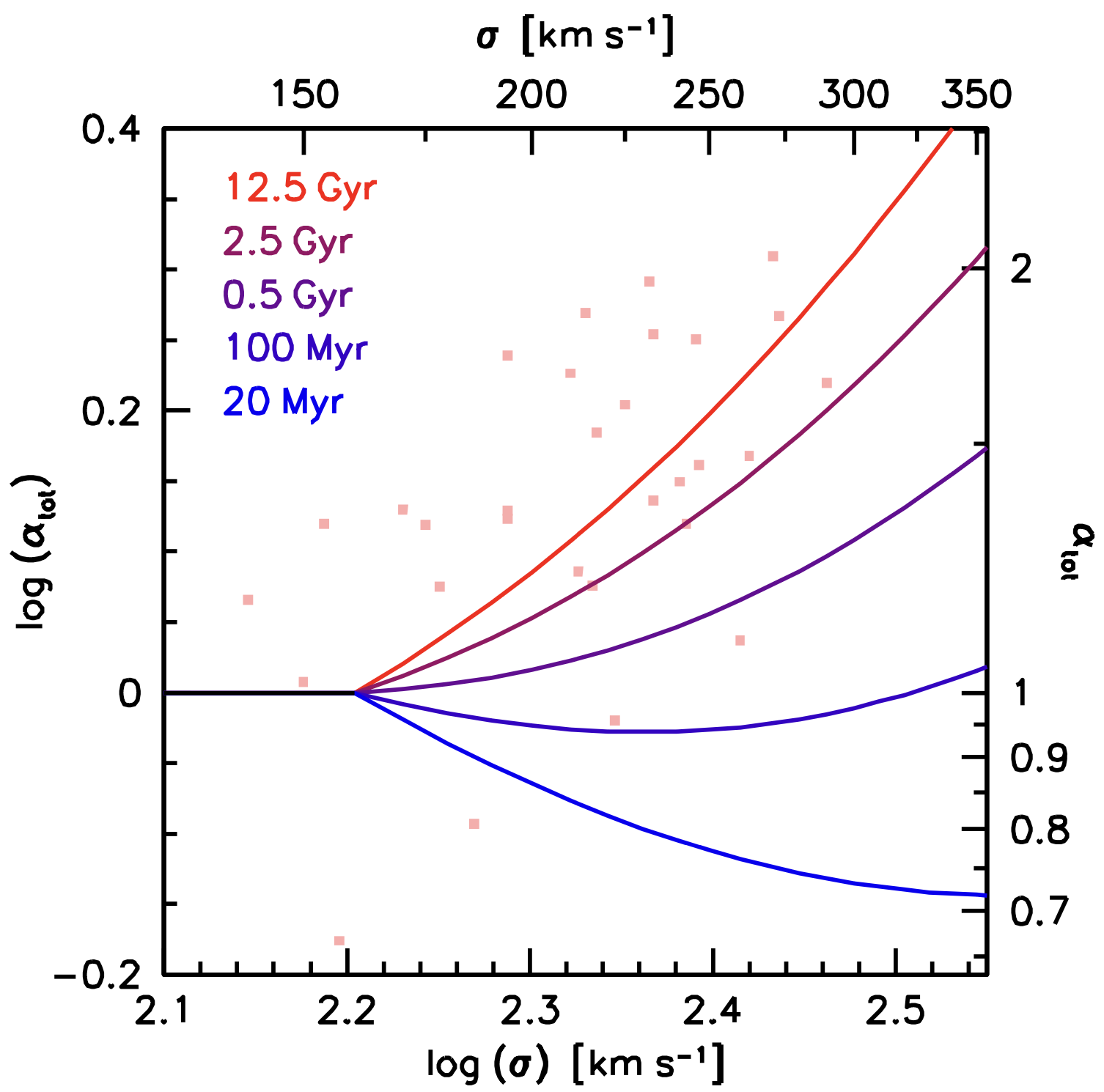}
  \end{center}
\vspace{-0.2cm}
    \caption{
Relation between the IMF parameter $\alpha$ and velocity dispersion, for the generalized concordance
IMF of Eq.\ \ref{gen.eq}. Data for the central regions of
nearby early-type galaxies from \citet{conroy:imf12} are shown in light red.
By construction, the model for an age of 12.5\,Gyr fits these data well. The other lines show the behavior
at younger ages.
}
\label{sigalpha.fig}
\end{figure}

We can take a first step toward such a description by assuming that
the form of the IMF varies smoothly as a function of velocity
dispersion.  Defining $\sigma_{x} \equiv \sigma/x$\,\kms, a bisector
fit to the \citet{conroy:imf12} data gives
\begin{equation}
\log \alpha_{\rm tot} \approx
0.22 + 1.16 \log \sigma_{250},
\label{alpha.eq}
\end{equation}
for $\sigma \gtrsim 160$\,\kms.
The concordance IMF has $\alpha_{\rm tot}=1.95$ for an age of 12.5\,Gyr, and is therefore appropriate for
galaxies with $\sigma \approx 290$\,\kms. Using Eq.\ \ref{alpha.eq}, and assuming
that galaxies with $\sigma\lesssim 160$\,\kms\ have a standard \citet{kroupa:01} IMF,
we obtain the following relations between the power law slopes of the IMF and the velocity dispersion:
\begin{eqnarray}
\gamma_1 &\approx& 1.3 + 4.3 \log \sigma_{160};\nonumber\\
\label{gen.eq}
\gamma_2 &\approx& 2.3 - 1.2 \log \sigma_{160};\\
\gamma_3 &\approx& 2.3 - 1.7 \log \sigma_{160},\nonumber
\end{eqnarray}
for $\sigma \gtrsim 160$\,\kms. For lower
velocity dispersions we simply have $\gamma_1 = 1.3$, $\gamma_2 = \gamma_3 = 2.3$.
The IMF parameter $\alpha$ of this generalized concordance IMF is shown in Fig.\ \ref{sigalpha.fig}.

For reference, measurements from \citet{conroy:imf12} are overplotted
as light red dots. These data are well described by the 12.5\,Gyr old
model, as expected. The other lines are the implied relations for
younger ages.  If galaxies primarily grow inside-out their velocity
dispersions do not change very much with time. This means that these
same relations can also be used at high redshift, as long as $\sigma$
can be measured, either directly from spectra or estimated from sizes
and masses \citep[e.g.,][]{bezanson:11}.

\section{Discussion}

Our analysis shows that it is possible, with a single form of the IMF,
to have $M/L$ ratios that are $\lesssim 0.8\times$ that of a Milky Way
IMF at young ages and $\approx 2\times$ that of a Milky Way IMF at old
ages.  The proposed concordance IMF with $\gamma_1=2.4$,
$\gamma_2=2.0$, and $\gamma_3=1.85$ ties together observations of the
cores of massive galaxies at $z=0$, dynamical studies of quiescent
galaxies at $2<z<5$, and recent JWST observations of unexpectedly
dense and massive galaxies at $z>7$.  At the highest redshifts the
mass-increasing effect of a steep low mass slope is more than
compensated by the luminosity-increasing effect of a flat high mass
slope.  The concordance IMF, and the generalized form (Eq.\
\ref{gen.eq}), can be implemented in a straightforward way, as
$\gamma_1$, $\gamma_2$, and $\gamma_3$ are defined over the same mass
intervals as the \citet{kroupa:01} IMF (which has $\gamma_1=1.3$, and
$\gamma_2=\gamma_3=2.3$).

We note that our results are insensitive to progenitor bias, that is,
the fact that samples of massive galaxies
at high redshift always represent only a subset
of all progenitors of such galaxies today
\citep{franxvandokkum:96,dokkumprog:01,leja:13,torrey:17}. The concordance IMF produces a
straightforward age dependence of the IMF parameter $\alpha$,
irrespective of the redshift, history, or future of a particular galaxy.
In this context it is encouraging that the concordance IMF
reproduces the low $M_{\rm dyn}/M_*$ ratios,
consistent with $\log\alpha \lesssim 0$, that have been reported for
massive, compact star forming galaxies at $z=2-3$
\citep{barro:14,barro:17,dokkum:15compact}.

Compared to standard Kroupa and Chabrier IMFs, the changes are modest
for the youngest ages --- for an age of 100\,Myr we have
$\log \alpha \approx 0$, independent of velocity dispersion
 --- and a conclusion from this study is that a
Milky Way-like IMF produces reasonable masses for massive
JWST-discovered galaxies. The effects can be more complex than is
indicated in Figs.\ 4 and 5, however, as changing the IMF not only
changes the luminosities but also the colors. To assess the magnitude
of this effect the three Balmer break galaxies at $z=7-8$ of
\citet{wang:24balmer} were re-fit with the concordance IMF, a standard
Kroupa IMF, and a Chabrier IMF.  The differences in $\log M_*$ between
the concordance IMF and the standard IMFs are in the range $-0.3$ to
$-0.2$, broadly consistent with expectations (B.\ Wang, priv.\ comm.).

We are not the first to explain observations of massive early-type
galaxies with IMFs that are simultaneously bottom-heavy and top-heavy,
although the precise slopes and mass ranges vary.  \cite{arrigoni:10}
find that a mildly flat slope at $>1$\,\msun, with
$\gamma_3 \approx 2.15$, can explain the slope of the
[Mg/Fe]\,--\,$\sigma$ relation for early-type galaxies.  In the same
vein, \citet{denbrok:24} find a steep low mass slope and a flat high
mass slope from a combined analysis of gravity-sensitive spectral
features and stellar abundances.  \citet{sande:15} introduce an IMF
with a slope of $\approx 1.4$ in the mass range $1-4$\,\msun\ (and 2.3
at lower masses) to explain the relation between color and $M/L$ ratio
for a sample of massive quiescent galaxies out to $z\sim 2$.
\citet{peacock:17} and \citet{coulter:17} favor these kinds of IMFs to
account for the relatively large numbers of X-ray binaries (i.e.,
stellar remnants) in elliptical galaxies, while maintaining a steep
low mass slope. Specifically, the \citet{coulter:17} form has a slope
of 3.8 at $<0.5$\,\msun\ and a slope of 2.0 for all masses
$>0.5$\,\msun.  In summary, mildly flatter slopes beyond $1$\,\msun\
are not only allowed by previous work but appear to be favored, for a
variety of reasons.

% check out
% Gunawardhana, M. L. P
% high sfr -> flatter imf
% maybe depends on SFR surface density
% (Weidner et al. 2011; Gunawardhana et al. 2011; Zhang et al. 2018)

%not much wiggle room in slope
%flatter imfs than 1.85 remnants overwhelm mass at late times
%example: img with slope = x produces y
%also applies to imfs proposed by steinhardt, woodrum - these lead to even lower m/l ratios
%at z=7, but produce huge remnant populations - quantify.
%Other parameter is low mass cutoff - can lower the stellar mass , leaving more room for
%ne of the best spectra - conroy shows tons of dwarfs
%on the other hand, snells-0

%color - m/l ratio: sande:15
%arrigoni:10: chemical evolution, explain Mg/Fe ratios.
%Another: den Brok et al 2024, based on chemical abundances arxiv:2404.03939

Our results can be tested in various ways. Thanks to {\it JWST} the
number of high redshift quiescent galaxies with dynamical masses is
expected to increase rapidly in the coming years, and it will be
interesting to see if $M_{\rm dyn}/M_*$ ratios show a similar trend as
$\alpha$ does in Fig.\ 4. Care must be taken, as many high redshift
galaxies are likely supported by rotation, complicating the interpretation
of observed velocity dispersions \citep[see, e.g.,][]{vanderwel:11,newman:18,bezanson:18}.
Special objects, such as the Einstein ring
JWST-ER1 \citep{dokkum:24lens,mercier:24}, provide unique information
that can be used to calibrate larger samples.\footnote{The reported
  values of $\log(\alpha)$ currently range between 0 and $0.3$ for
  this object, depending on the assumed redshift of the ring. A Cycle
  3 JWST program (GO-5883; PI R.\ Gavazzi) will resolve this
  question.}  Direct measurements of low mass stars at significant
redshifts would be even more constraining: as shown by the dotted line
in Fig.\ 4, the mass excess in living stars should show a similar
trend as the total mass excess, and the interpretation of the data is
more straightforward (as it is independent of the mass in stellar
remnants, dark matter, and gas).\footnote{The Cycle 3 JWST program
  GO-5629 (PI: M.\ Kriek) aims to perform this measurement at
  $z\sim 0.7$.}  Stellar population fits with non-parametric forms of
the IMF may be able to differentiate between a constant steep slope
below 1\,\msun\ and a gradually increasing one. Encouragingly, there
is evidence for such an increasing slope in NGC\,1407, the only galaxy
that has been analyzed this way so far \citep{conroy:17}. Lastly, an
important caveat is that the steep low mass slope is currently based
on a single methodology, the measurement of gravity-sensitive
absorption lines in the integrated spectra of early-type galaxies. All
other observations could also be explained by a more straightforward
top-heavy, or bottom-light, IMF, as stellar remnants would take the
place of low mass stars in producing the high $M/L$ ratios of
present-day massive galaxies. Independent tests of the low mass IMF,
such as the one we introduced in \citet{dokkum:21chromo}, would be
highly valuable.\footnote{In yet another Cycle 3 JWST program, GO-4757
  (PI: P.\ van Dokkum), we plan to use the strength of near-IR H$_2$O
  features to constrain the low mass IMF.}

The physical origin of IMFs that are simultaneously bottom-heavy and
top-heavy is unclear.  The behavior of the IMF at low masses has long
been linked to the Jeans mass and its dependence on temperature,
density, and Mach number.  In extreme environments the low-mass IMF
can be supressed due to the $T^{3/2}$ temperature dependence of the
Jeans mass \citep{larson:98,krumholz:06,bate:09,bate:23}, or enhanced
if the $\rho^{-1/2}$ density dependence or Mach number dependence
dominates \citep[e.g.,][]{hopkins:13,chabrier:14,tanvir:22,tanvir:24},
but it is difficult to do both at the same time \citep[see,
e.g.,][]{bate:23}. A different process may operate at high
masses. Models based on Press-Schecter-like arguments
applied to turbulent molecular clouds \citep{hennnebelle:08,
hopkins:13,chabrier:14} directly link the high-mass IMF slope to the
spectral index of the turbulent velocity power spectrum, with steeper
spectra giving rise to shallower high-mass IMF slopes.  
These models have been tested and refined
with hydrodynamical simulations \citep{nam:21}, although
it remains to be
seen whether the required modifications to the spectral index occur in
realistic environments. To summarize, $\gamma_1$ may be largely determined by
the Jeans mass and $\gamma_3$ by the index of the
turbulent power spectrum, but more work is needed to test this.

%Such models
%may provide a natural framework to simultaneously produce bottom-heavy and
%top-heavy IMFs by invoking turbulence as the key process working at
%both ends. 

It may also be that the observed IMF is a combination of star formation in
different local environments within early galaxies, with the densest
regions producing the steep low mass slope and the regions that are
most exposed to radiation producing the flat high mass slope. 
{\it JWST} results have shown that many early galaxies have extremely
high central surface densities \citep{baggen:23,carnall:24}, lending
some observational support to this idea.  Such hybrid IMFs have been
developed previously in the framework of the `integrated galactic IMF'
\citep[IGIMF;][]{kroupaweidner:03}, and the IMF that \citet{jevrabkova:18},
\citet{yan:19,yan:21},
and \citet{haslbauer:24} propose for high metallicity, high star
formation rate environments is qualitatively quite similar to the
concordance IMF. \citet{fontanot:18}, working from
similar concepts as \citet{kroupaweidner:03}, introduce
physically-motivated IMFs that have a break at 1\,\msun\ with a steep
slope at low masses and a shallow slope at high masses. As shown in
\citet{fontanot:23}, these IMFs reproduce the $M/L$ ratios of
present-day ellipticals well, although they underpredict the number of
low mass stars (the mass excess with respect to the Milky Way IMF is
largely in the form of remnants, not low mass stars, in these models).

Several of the questions that are raised in the preceding paragraphs
could be addressed if local analogs could
be studied. Returning to the evidence for top-heavy IMFs discussed in the
Introduction, the high mass slope in 30 Doradus is $1.90^{+0.37}_{-0.26}$
\citep{schneider:18},
similar to that of the concordance IMF. It would be very interesting ---
although observationally challenging; see, e.g., \citet{fahrion:24} --- to
determine if there is an upturn at low masses in this cluster. We note that if such analogs exist in the Local Group, it would 
imply that the observed relations with global galaxy
properties such as $\sigma$ are proxies for the preponderance of
particular local conditions within the galaxies.

As a final remark, we acknowledge the long history of appealing to IMF
variation as a means of explaining real or imagined observational
tensions, and that in such previous cases a more prosaic explanation
was usually found.  We therefore suggest adopting the concordance IMF
with caution, for now.

\vspace{0.5cm}

\noindent
We thank Mariska Kriek and Bingjie Wang for their comments on the
manuscript.  Bingjie Wang also fitted the galaxies in her paper with
the concordance IMF. We are grateful to the referee, Mark Krumholz,
for insightful comments that improved the paper.
Finally, we thank the organizers of the 2011 Galaxy
Formation conference in Durham, as this was when we first considered
`ski slope' IMFs as a potential way to reconcile seemingly
contradictory results in the literature.

\bibliography{master_0811}{}
\bibliographystyle{aasjournal}

\end{document}